\documentclass[prodmode,acmtoit]{acmsmall} 
\usepackage[english]{babel}
 \usepackage{listings}
 \usepackage{float}
 \usepackage{longtable}
 \usepackage{enumerate}

\usepackage[ruled]{algorithm2e}
\usepackage{booktabs}

\SetAlFnt{\small}
\SetAlCapFnt{\small}
\SetAlCapNameFnt{\small}
\SetAlCapHSkip{0pt}
\IncMargin{-\parindent}

\acmVolume{9}
\acmNumber{4}
\acmArticle{39}
\acmYear{2017}
\acmMonth{5}

\doi{0000001.0000001}

\issn{1234-56789}

\begin{document}

\markboth{Bellini et al.}{Quantify resilience enhancement of UTS through exploiting Connected Community and Internet of Everything emerging technologies}

\title{Quantify resilience enhancement of UTS through exploiting Connected Community and Internet of Everything emerging technologies}
\author{Emanuele Bellini
\affil{University of Florence}
Paolo Ceravolo
\affil{University of Milan}
Paolo Nesi
\affil{University of Florence}
}


\begin{abstract}
This work aims at investigating and quantifying the Urban Transport System (UTS) resilience enhancement enabled by the adoption of emerging technology such as Internet of Everything (IoE) and the new trend of the Connected Community (CC). A conceptual extension of Functional Resonance Analysis Method (FRAM) and its formalization have been proposed and used to model UTS complexity.  The scope is to identify the system functions and their interdependencies with a particular focus on those that have a relation and impact on people and communities. Network analysis techniques have been applied to the FRAM model to identify and estimate the most critical community-related functions.
The notion of {\em Variability Rate} (VR) has been defined as the amount of output variability generated by an upstream function that can be tolerated/absorbed by a downstream function, without significantly increasing of its subsequent output variability.
A fuzzy based quantification of the VR on expert judgment has been developed when quantitative data are not available. Our approach has been applied to a critical scenario (water bomb/flash flooding) considering two cases: when UTS has CC and IoE implemented or not.
The results show a remarkable VR enhancement if CC and IoE are deployed.

\end{abstract}

%
%


%
%



\acmformat{TBD}

\begin{bottomstuff}
This work is supported by the H2020-RESOLUTE project
\end{bottomstuff}

\maketitle

\section{Introduction}

The effectiveness of the current risk and efficiency-based approaches in complex socio-technical systems safety and security management is affected by  their weakness in addressing the so-called ``unknown unknowns" \cite{park2013integrating}. This is caused by the continuous increment of the complexity of the systems and the emergent and unpredictable conditions such as climate change or man-made sabotages. According to \cite{field2012managing},  the climatic extremes may intensify or become more frequent in regions that are not used to cope with such events. Moreover, so far no scientific method is available to precisely predict the long-term evolution and spatial distribution of critical events, nor  the impacts on society's critical infrastructures. \\

Complex socio-technical systems cannot be managed under the assumption that accidents are produced by an uncontrolled and undesired release or transfer of energy between technical components \cite{leveson} and the large number of human/social, organisational and technical aspects, together with their fast pace changing behavior, imposes serious limitations on the ability to fully understand and monitor system operations. 
Therefore, complex socio-technical systems are today underspecified by nature \cite{Wilson2006} and a certain level of epistemic and aleatory uncertainty must be taken into account as a contribution to the critical events. 

As \cite{owens} pointed out, accidents within complex environments tend to be the result of unpredicted interactions, rather than single failures of human or technical components. This produces unexpected cascade effects, which could rapidly reach unacceptable proportions.  In order to face these unknown elements, building resilience becomes the best decision for socio-technical systems as the Urban Transport Systems (UTS)\cite{linkovnature}. There are many definitions and interpretations about resilience\footnote{ RESOLUTE D2.2 Sate of the art}. In the context of RESOLUTE\footnote{ RESOLUTE is an EC funded research project - http://www.resolute-eu.org}, it refers to the capability of a system of continuously adapting to its operational environment in the pursuit of its intentions/purposes. Thus resilience can be defined as the ability of a system to sustain required operations in both expected and unexpected conditions  by adjusting its functioning prior to, during, or following changes. 
According to the resilience engineering field, the potential for resilience to emerge from system performance may be assessed based on the {\em four resilience cornerstones} \cite{hollangel2011b}, \cite{Hollnagel2015}:
\begin{enumerate}[a)] 
\item Knowing what to do - corresponds to the ability to respond to disruptions by adjusting system performance to changing conditions.
\item Knowing what to look for - corresponds to the ability to monitor both the system and the environment.
\item Knowing what to expect - corresponds to the ability to anticipate opportunities for changes in the system and identify sources of disruption and pressure and their consequences for system operation.
\item Knowing what has happened - corresponds to the ability to learn from past experiences of both successes and failures.
\end{enumerate}

In other words, the essence of resilience is the ability of the system to recognize when variability in its performance is unanticipated and fall beyond the usual range, and to dampen such variability through continuos adaptation. 
In order to cope with such a variability and to respond to different and possibly conflicting local operational needs, the limited resources of the system (humans, technologies and organization) should be managed  and exploited effectively to achieve the right system synchronization and coordination level needed to ensure successful operation. However, it is clear that the variability and uncertainty need to be considered as intrinsic characteristics of complex socio-technical systems \cite{hollnagel2008fram}. 

 According to this perspective, the new global trend of Internet of Everything (IoE) in general and the Connected Community (CC) in particular, can be exploited as resources of the socio-technical system to enhance its adaptive capacity and thus the resilience of the UTS dampening unwanted variability. The IoE can be considered a natural development of the IoT concept. In fact, while "Things" are related to connect physical-first objects, IoE extends this view comprising the following four key elements including all sorts of possible connections:
\begin{enumerate}[a)] 
\item People: Considered as end-nodes connected across the Internet to share knowledge, information, opinions, decisions, behaviors and activities. 
\item Things: Physical sensors, devices, actuators and other items generating data or receiving information from other sources. 
\item Data: Raw data analyzed and processed into useful information to enable intelligent decisions and control mechanisms (e.g., Human behaviors on the ground). 
\item Processes: Leveraging connectivity among data, things and people to add value. 
\end{enumerate}
Thus IoE establishes an end-to-end ecosystem of connectivity where people with their relationships, social collaborations and grouping dynamics represent an integral part. 
In particular, according to \cite{bpcc}, Connected Communities are characterised, among the others,  by weak ties that can symbolise a range of potential relationships among community members.  These relations range from tight, long-lasting and static  to temporary, real-time and dynamic relationships of different durations to location-specific connections. In fact, communities can be established on the base of same interests, skills or because of being at the same place and time in relation to some adverse event. Such a CC characteristics can be exploited to shift from a public awareness approach to one of community-individual safety altering the traditional top-down ``command and control" relationships with the population. In fact, in RESOLUTE, the community is seen as an active participant to build the system resilience, rather than a passive recipient of services. Hence, the IoE and CC, if properly exploited, can be considered as means to achieve resilience in UTS because they could:   
\begin{enumerate}[a)] 
\item  enhance the monitoring  and control capability, improving the granularity and breadth of knowledge and awareness about the system status and dynamics continuously collecting Big Data from heterogeneous data sources/streams and sensors  as people GPS position, concentration, behaviors and sentiment through smart devices and social networks (User Generated Data),  Open Data, data from environmental sensors (e.g.,  traffic flows, hydrometry, air pollution, underpasses water level), mobile cell data, wifi access points, and real-time reports such as weather forecast, and so forth \cite{dms};
\item enhance the responding capability by providing detailed and timely information to authorities on one side, and to delivering personalised, real-time, context-aware, and ubiquitous advice to the community exploiting  technologies such as IoE, Fast Wireless Connections (free wifi, 3G/4G), LoRaWAN, Smart Mobile Devices,  Big Data Analytics, Semantic Computing, etc., that are crucial for augmenting situation awareness and enhancing  decision making; 
\item enhance the learning capability applying advanced analysis on Big Data (e.g., deep learning, data analysis and prediction, sentiment analysis)  to extract knowledge; 
\item enhance then anticipation capability continuously supporting the assessment of vulnerability and identifying when the system operates nearer to safety boundaries, predict behaviors and event dynamics, support evidence-based decisions at strategic, tactic and operation level moving ahead respect the current practices based on pre-simulated emergency scenarios \cite{Woltjer}. 
\end{enumerate}
Unfortunately, even if several initiatives are ongoing at international levels such as the political UNISDR Sendai Framework\footnote{UNISDR - http://www.unisdr.org/we/coordinate/sendai-framework}, cities and local communities are slow in becoming smart and resilient because of several factors such as budget restriction, cultural gaps, and  by the difficulties to quantify the benefits for the community (e.g. Social Return Of Investment). In fact, because of resource scarcity, a priority rank for infrastructure improvement actions tends to be based on political opportunity or heuristics instead of a quantitative evaluation of the benefit of the system as a whole. \\

To this end, the present article aims at demonstrating and quantifying the enhancement of UTS resilience obtained with the exploitation of IoE and CC as enabling technologies capable of significantly increasing the variability dampening  capacity of those functions in UTS related to the human/social aspects.    \\
Defining a method for variability quantification enables also the development of the so-called Big KID--driven Decision Support System\footnote{KID stays for  {\em knowledge, information, data}.}. A Decision Support System (DSS) \cite{dorasamy2013knowledge}, \cite{tsekourakis2012decision}, \cite{suarez2013improving}, \cite{bartolozzi2015smart} is a computer-based information system that supports organizational decision-making activities. The objective of a DSS is to provide evidence for making decisions for a problem by compounding experts' experiences and data and analyzing them in an intelligent and fast way a human cannot do in reasonable time.

Hence to achieve research intent, the work has been organised in the following 3 steps:
\begin{enumerate}[a)] 
 \item The complex socio-technical system (e.g., the UTS) and the role of the CC in daily operations has been analysed through the Functional Resonance Analysis Method \cite{hollnagel2008fram} perspective. The FRAM is a method to analyse how the activities daily take places daily in the complex system and introduces powerful concepts as functional variability, dampening, adaptive capacity, functional resonance, etc.as well as a specific notation to model the systems that is described in section 3. However the lack of an effective formalization of FRAM prevents to carry out the quantitative assessment of the impact of the IoE and Connected Communities in the UTS resilience building. 
\item A new formalization and a method to quantify the FRAM functional variability and the dampening capacity has been defined 
\item The new method to quantify FRAM has been tested in a case study comparing the potential variability in UTS with or without IoE and Connected Community exploitation. The benefit in terms of dampening capacity has been quantified.
\end{enumerate}

This article is organised as follow: in Section 2,  the role of people in the context of UTS  is presented; in Section 3, we introduce the background work about the FRAM based Critical Infrastructure Reference Model published in the RESOLUTE European Resilience Management Guidelines; in Section 4, a new methodology to quantify the variability in FRAM is proposed; in Section 5, an example of the application of the methods to govern CC behavior during emergency exploiting IoE technologies is represented; in Section 6, conclusions and next steps are discussed.

\section{Urban Transport System and people}

 In the UTS, operations have developed a prominent safety and business critical nature, in view of which current practices have shown the evidence of important limitations in terms of resilience management. Hence, enhancing resilience in UTS is considered imperative for two main reasons: 
\begin{enumerate}[a)] 
\item such systems provide essential support to every socio-economic activity and rescue,  and
\item the paths that convey people, goods and information, are the same through which risks are propagated and resource are provided  \cite{esrel}.
\end{enumerate}
Unfortunately, even if the UTS plays a critical role in the society, there is a general tendency to leave out from resilience strategy implementation, crucial aspects such as the coordination and synchronisation amongst several system functions and elements as the community preparedness and behavior and the need to account for a wide range of unknown scenarios and context dependent factors. In fact, humans do not have the time, the mental resources or the capability to be aware of every problem at the same time. They devote their energy to problems that involve them and for which they can make a difference - J. E. Grunig quoted in Leffler \cite{MerrilLef}. Thus the community members need to be enabled and engaged as an active participant in his/her own safety developing a self-resilience attitude. For instance, in a situation where the number of options to escape from a hazard are limited or absent because of the presence of constraints as bridges or tunnels, CC members can help each other or receive valuable information from the first responders to adopt specific behaviors to mitigate the impact of the event. 
This requires new technologies, new skills and new approaches to enable users in being connected anytime and everywhere to provide and receive lifesaver information and adapt their behavior accordingly.
In particular,  every aspect of the human factor (behavior, attitude, belief, sentiment, skill, heuristics, etc.) needs to be considered in a critical infrastructure like UTS, as key elements for resilience building, going beyond the engineering and operational approaches that tend to be focused on technologies and procedures.
Moreover, the different nature of UTS users (i.e. cars, motorcycles bicycles, pedestrians, among others) and the wide diversity of purposes encompassed within urban transport, tend to generate highly dynamic interdependencies, both within the private transport system and with public transport. It is clear that, governing/directing CC behaviors during the UTS usage, is the basis of solutions towards enhanced resilience discussed in this article. \\

Beyond the aspects of system complexity, the global scenario of resource scarcity and changes is also put forward as a cause for many of the serious safety and security threats currently faced by societies. In  \cite{Boin},  such threats are distinguished from {\em routine emergencies} such as fires and traffic accidents, and characterise them as {\em low-chance}, {\em high-impact} events that can compromise life sustaining systems and require governmental intervention under high uncertainty conditions. 
Both are the circumstances in which resilience is highlighted as a possible solution for the sustainability, reliability and safety of systems \cite{Boin} and \cite{jackson}. In fact, in resilience engineering field, there is not a difference between routine and big events, what change is the amplitude of the functional variability and thus the possibility of emerging resonance effect among the system functions.

Such an IoE enabled data-driven approach provides the means to assess the levels of criticality at evidence/quantitative level, while seeking to enable the capabilities of the complex system to take the appropriate decision at strategic, tactical and operational levels \cite{bellini2014km4city}.\\

\section{Understanding UTS Behavior through FRAM}

\begin{figure}[!ht]
\centering
    \includegraphics[type=png,ext=.png,read=.png,width=12cm]  
{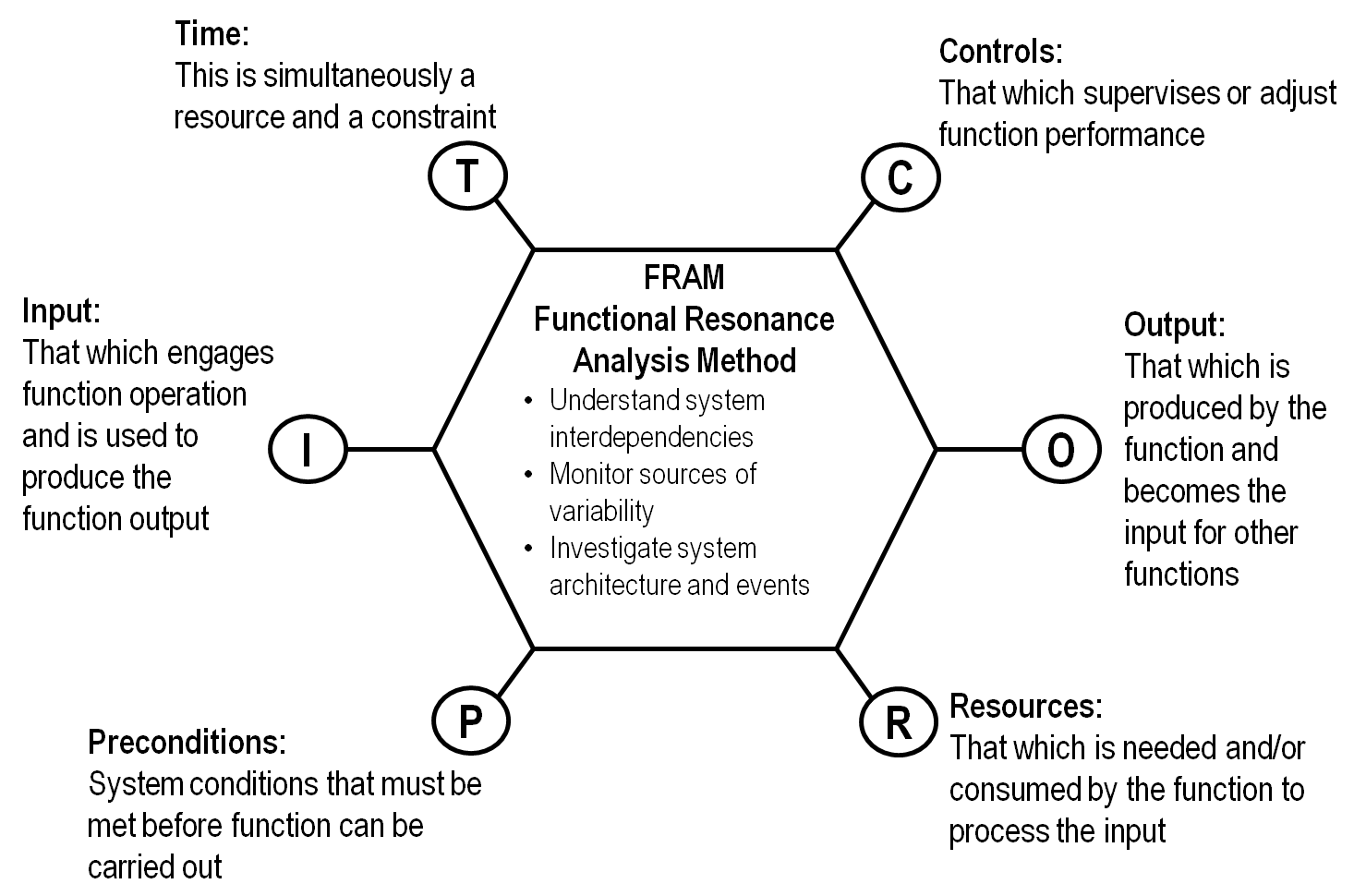}
    \caption{Functional unit of FRAM (adapted from Hollnagel, 2008).}
    \label{fig:FunctionalUnitFRAM} 
\end{figure}

The system analysis is based on the Critical Infrastructure reference model defined in the European Resilience Management Guidelines\footnote{E. Bellini, P. Ferreira, and E. Gaitanidou. 2016a. European Resilience Management Guidelines (h2020 RESOLUTE project ed.) }  where the human  aspect and the community management is included in the system description. The FRAM \cite{hollnagel2008fram} was used to support system analysis, aiming to identify interdependencies and system emergent behaviors potentially relevant for resilience. The FRAM approach is essentially a system-modelling tool that focuses on system interdependencies, their dynamics and complexity. It is grounded on Resilience Engineering principles and provides a fundamental support to such ends by supporting systems understanding. In particular, a system is considered a set of coupled or mutually dependent functions.
FRAM is particularly relevant in describing nonlinear systems and the overall rule is to try achieving a description of the normal activities performed by the socio-technical system involving stakeholders in its definition. 
A FRAM model is illustrated in Figure \ref{fig:FunctionalUnitFRAM} were a {\em function} $F$ is composed by a label (usually a verb) representing the action of the function and by  six {\em aspects} $A$. 

Hollnagel in \cite{hollnagel2008fram} defines the six {\em aspects} in the following terms: 
\begin{enumerate}
\item \textbf{Input}: that which the function processes or transforms or that which starts the function.
\item \textbf{Preconditions}:  that must exist before a function can be executed.
\item \textbf{Resources}: that which the function needs or consumes to produce the output.
\item \textbf{Time}: as temporal constraints affecting the function (with regard to starting time, finishing time, or duration).
\item \textbf{Control}: how the function is monitored or controlled.
\item \textbf{Output}: is the result of the function, either a specific output or product or a state change.
\end{enumerate}

It is important to notice that the first five aspects (Input, Preconditions, Resources, Time, Control) are acting as {\em inputs} while the function {\em outputs} are represented only by the Output aspect. 
The characterisation of the functions, in terms of the six aspects, contains the potential couplings among functions. In fact, the input aspects of a downstream function can receive a qualified output from upstream functions. Such qualified output is a {\em relationship} $R$ labeled with a textual definition and representing the tangible or intangible outcome of the function of origin towards the function of destination. In fact, each output can be the input of another function.  \\

\begin{figure}[!ht]
\centering    \includegraphics[type=png,ext=.png,read=.png,width=15cm]
{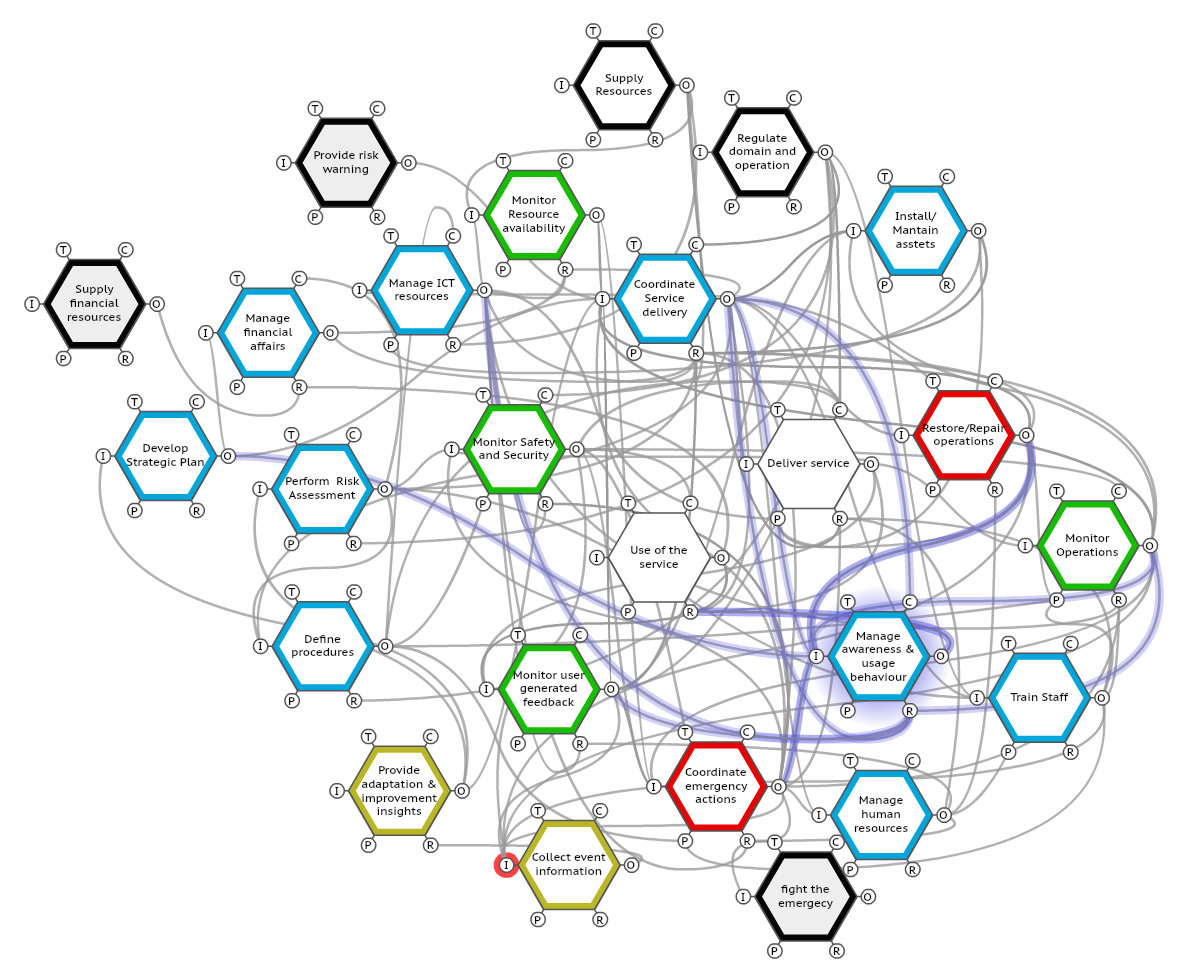}
\caption{RESOLUTE UTS Model from European Resilience Management Guidelines\footnote{E. Bellini, P. Ferreira, and E. Gaitanidou. 2016a. European Resilience Management Guidelines (h2020 resolute project ed.).} }
    \label{fig:UTSmod}
\end{figure}

On the basis of FRAM approach, Figure \ref{fig:UTSmod} reports the desired functions and interdependencies that a UTS needs to implements  to be resilient\footnote{E. Bellini, P. Ferreira, and E. Gaitanidou. 2016a. European Resilience Management Guidelines (h2020 reso- lute project ed.).}. In Appendix Table \ref{tab:functions} lists the functions composing the model with their relationships.

According to the ERMG, the human/social components of the socio-technical system are addressed, at least, by the following functions: {\em Use of the service}, {\em Manage awareness and user behavior}, {\em Monitor user generated feedback} which is defined as follow: 
\begin{itemize}
\item Use of the service: This function represents the actual usage of the service like driving a car, goods movement and delivery, taking taxi, bus or metro, walking etc. in the UTS.
\item Manage awareness \& user behavior: as providers of fundamental public services, critical infrastructures tend to be significantly exposed to individual and collective behaviors, in many cases, not just the service end-users, but also of the wider public. Recent technological developments, in particular in relation to ICTs, offer a great potential for the enhancement of interactions with the public and the use of this potential towards an increased effectiveness in managing and deploying operational adjustments to various relevant events and circumstances.
\item Monitor user-generated feedback: monitor feedbacks about service usage on a wide range of parameters and produce fundamental support to the deployment of operational adjustments. This function deals with the need for an integrated approach to the assessment of user generated feedback, mainly by placing this data and information in the context of operational monitoring. 
\end{itemize}

\subsection{Extending performance variability concept}
The FRAM approach is based on the principle of equivalence of successes and failures and the principle of approximate adjustments thus performance is therefore in practice always variable.
As thoroughly explained by Hollnagel \cite{hollnagel2008fram} {\em performance variability}, i.e., the range of result in a function or an overall system's performance, is highly dependent on the variability of the conditions under which the system/function is performing. 

Starting from the generic six main sources of human and organisational performance variability defined in \cite{hollnagel2008fram}, it is possible to identify the following in the UTS domain:
\begin{itemize}
\item Fundamental human physiological and/or psychological characteristics as  driving fatigue, vigilance, attention, risk perception  of UTS users, etc.
\item	Pervasive higher  level psychological phenomena like adaptability as taking decisions within UTS knowledge uncertainty.
\item	Organisational conditions and requirements, as the need to meet external demands, stretching resources, substituting goals, etc.
\item Social or team psychological factors, such as meeting expectations of oneself or of colleagues, complying with group working standards, etc.
\item Context variability: roads conditions are too hot, too noisy, too crowded, etc.
\item Environment variability induced by the unpredictability of the domain, e.g., weather conditions, technical problems, etc.
\end{itemize}

According to FRAM, performance variability is assessed through the eleven Common Performance Conditions (CPC), verifying if their performances are stable or variable but adequate, stable or variable but inadequate, or unpredictable \cite{hollnagel2012}. The variability about the way a function is carried out may show itself by the variability of its {\em output}. Since the results generated in {\em output} by a function can affect other aspects, namely {\em input}, {\em precondition}, {\em resource}, {\em time}, or {\em control}, of one or more downstream functions. Note that, the range of behaviors and effects captured by the {\em output} element is very broad and includes any exchange of matter, energy, or information. The {\em output} can be seen as representing a change of state in the system or in one or more aspects of downstream functions. But the {\em output} can also represent a decision or a signal that starts a downstream function. Moreover, in complex and non-linear systems predicting the specific outcomes of a function can be hard or unmeaning. For this reason, the literature concentrated on characterizing function variability in term of performances. The following dimensions, that are a combination of what proposed in the FRAM Glossary\footnote{ FRAM Glossary - http://functionalresonance.com/a-fram-glossary.html}, are considered relevant in UTS:
\begin{itemize}
\item Timing: too early, on time, too late, not at all. 
\item Duration: too little, too much, right duration.
\item Distance: too close, too far, right distance.
\item Magnitude: too strong, too weak, right magnitude.
\item Speed: too fast, too slow, right speed.
\item Force/power/pressure: too high, too low, right force. 
\item Precision: precise, imprecise, right precision. 
\item Volume: too much, too little, right volume.
\item Costs: cost effective, costly, too much expensive. 
\end{itemize}

In this work, we refer to the performance variability of a single function as {\em Function Performance Variability} (FPV). We also underline that the FPV of upstream functions may affect the FPV of downstream functions, and thereby lead to non-linear effects called functional resonance. A resonance phenomenon in physics usually results in a significant increase in the amplitude of the oscillations, which corresponds to a considerable buildup of energy within the stressed system.
Similarly, functional resonance in the system emerges when the variability is spread through the interdependencies of the system functions causing the amplification of the effects until the system loses its capability to manage variability safely. \\

Even if the variability in function execution performance can be derived by the variability of its  {\em output},  the impact of such a variability over the system cannot be determined by observing the variability of output values only. In particular, we argue that it also depends on the variability acceptance supported by the function receiving inputs. In fact, the functional resonance effect is triggered by the rest of the variability of the upstream function output that is not absorbed  by the downstream function.
Moreover, the impact of variability is then intrinsically associated to relationships coupling outputs and inputs and can be expressed by the matching between output variability and input dumping capacity. This approach extends the current conceptualization of the FRAM providing a new concept useful for its formalization as discussed in the subsequent sections.

\section{Formalising FRAM}

In order to develop analytics over a FRAM representation, it is necessary to formalize the description of the target system. A similar attempt has been done by Cambrensis\footnote{http://www.cambrensis.org/wp-content/uploads/2012/05/Systemic-Interdependency-Modelling-GENSIM-0.1-docx.pdf} where FRAM  has been formalised with a dependency model based on Bayesian Belief Network to quantify functional variability. Such approach presents several advantages. It allows a rigorous formalization and the automatic update of all the relevant interdependencies among the FRAM functions, iteratively. Even if the variability propagation can be modeled weighting the arcs in the BBN,  the basic assumption behind this approach is that the entire variability of the upstream function output affects the downstream function performance. 
This means that any kind of adaptation capable of dampening input variability exhibited by a function, is not taken into account. 

For instance, the human resources usually engaged by a function could be incremented to absorb the arrival delay (variability) of an input in order to produce the expected output in due time\\

The instantiation of a FRAM is usually depicted as a directed graph where nodes represent the functions with their six aspects, taking the shape of a hexagon, and edges represent qualified relationships among functions by interconnecting two aspects. This representation is essentially oriented to human readability and does not offer any support to quantitative analysis.
In order to improve the current state of the art in executing quantitative analysis over a FRAM representation we are facing 4 objectives: 
\begin{itemize}
\item {\bf O1}. Representing dependencies among functions as well as qualified relationships. 
\item {\bf O2}. Representing the matching between performance variability and damping capacity intrinsic to relationships.  
\item {\bf O3}. Integrating quantitative and perception-based observations.  
\item {\bf O4}. Test our method with a contingency plan by comparing a scenario with IoE and CC deployed in UTS and a scenario without such technologies.
\end{itemize}

\subsection{Dependability analysis}
A typical analysis to be carried out on a FRAM model is related to the identification of the dependencies among functions.  The aim is to look at the couplings among functions in order to identify whether they will lead to unwanted outcomes that may compromise the process. 

In Systems Engineering and Risk Management \cite{thalmann2014integrated} Dependability is typically estimated by the number of originated errors, using metrics such as {\em Mean Time To Failure} \cite{delong2005dependability}. However, this approach imposes onerous observations and tests in a posteriori analysis and subjective observations in a priori analysis. For this reason, we propose to implement a quantitative analysis of the dependencies by representing the connections among functions and relationships using graph metrics \cite{hernandez2011classification}. As stated in {\bf O1} we do not want to limit our attention to functions.
Several relationships may be originated from a single function, thus to distinguish them and to measure their position in the graph we need to include them in the set of nodes considered by our analysis.

The most proper way to represent relationships $r \in R$ in a FRAM model is to use a quadruple $r = \{o, d, a, qn\}$, where $o \in F$ is the origin or upstream function, $d \in F$ the destination or downstream function, $a \in A$ specifies the FRAM aspects involved in the relationship, while $qn \in QNames$ is a qualified name for the relationship\footnote{For a definition of $QNames$ we refer the reader to \cite{weik2000computer}}. Note that the triple $\{o, d, a\}$ does not represent a sufficient condition for identifying a relationship as multiple links may interconnect two functions along the same aspect.   
In fact, the set of origin functions is included in the set of functions with relationships along the {\em output} aspect, or more formally: $O \in F \times A | a = output$. Similarly, the set of destination functions is included in the set of functions having {\em input}, {\em precondition}, {\em resource}, {\em control} or {\em time} as aspects, more formally: $D \in F \times A | a \in \{input, precondition, resource, control, time\}$.

These notions can be exploited to inspect, with an analytical perspective, the dependencies characterizing a FRAM, using a matrix to encode the graph structure resulting by the unification of the connections between functions and relationships.  
The simplest approach is to generate an {\em adjacency matrix} of a bipartite graph, i.e., a matrix $M$ that records the connections between two classes of objects, in our case $F$ and $R$, such as its element $m_{i,j} \in (0,1),$ is $1$ if $i$ and $j$ are related and $0$ if they are not. The properties of the matrix $M$ can be specified by stating that

$$ M = \left(\begin{array}{cc}0_{f,f} & B \\B^{T} & 0_{r,r}\end{array}\right),$$

where $B$ is an $F \cup R \times F \cup R$ matrix, $B^{T}$ is its transpose, and $0_{f,f}$ and $0_{r,r}$ represent the $F \times F$ and $R \times R$ zero matrices. 
Moreover, not all possible connections in $B$ and in $B^{T}$ are allowed because $F$ is the union of two disjoint sets $O$ and $D$; where, by definition, all $m_{i,j}$ with $i \in D$ and $j \in R$ or $i \in R$ and $j \in O$ are equal to $0$.
For example, if we know that the {\em output} of function $F13$ gives {\em input} to function $F14$ with a relationship named {\em User Behavior}, we can express the following by encoding two connections:  {\em F13 $\rightarrow$ F13:UserBehavior:F14:Input} and {\em F13:UserBehavior:F14:Input $\rightarrow$ F14}.
The resulting network has been weighted according to the importance of the relationship in the system. The weights assignment task has been conducted within the RESOLUTE project translating workshops and stakeholders interviews with Civil Protection and City Council managers, firefighters and citizens.\\

\begin{figure}[!h]
    \centering
    \includegraphics[type=png,ext=.png,read=.png,width=13.5cm]
{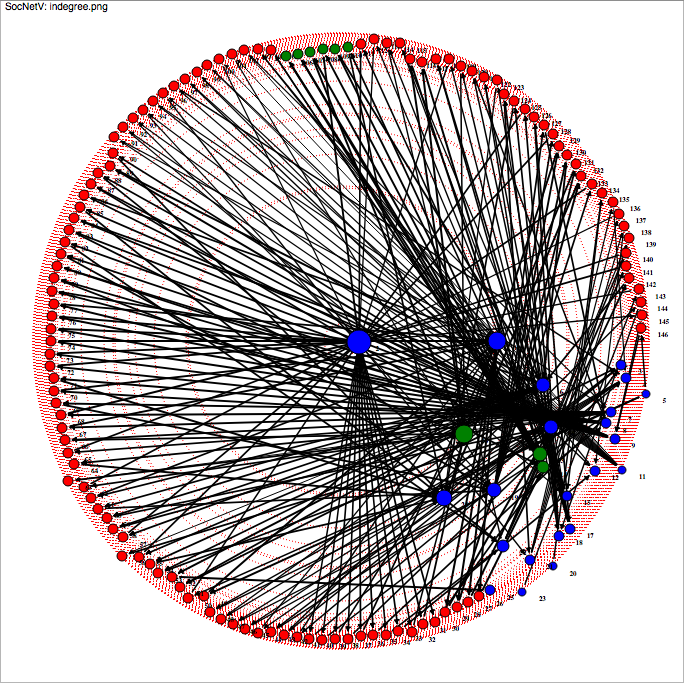}
    \caption{A graph $G_{f,r,m}$ ordering nodes in concentric ranges	based on their DP value.}
    \label{fig:indegree}
\end{figure}

The union of these connections provides us with a graph. A broad variety of measures to characterise graphs are exploited in several scientific domains \cite{scott2012social}. The ratio between  the number of vertices and edges reveals the {\em Sparsity} of a graph. The {\em Clustering Coefficient} is a measure of the degree to which nodes tend to cluster together.  {\em Node Centrality} gives a measure of how central in the overall graph a node is.
The method we adopted to quantify dependability of FRAM model is the Degree Prestige (DP) index \cite{Freeman}, a metric accounting the number of inward connections entering in a node. We claim this is the right choice, because in FRAM  a) the number of connections (explicit), b) the importance of connections (implicit), c) the direction of the relations (explicit) are critical elements characterizing the functions identified. Thus using a weighted and directed graph DP is the sum of weights of all connections ending at a given node, where nodes with higher DP are considered more prominent among others because they receive more inbound ``heavy'' connections. The largest the index is, the more prestigious/important the node is. \\
Clearly, other approaches can be followed. For instance, {\em Closeness Centrality} measures node centrality by considering the geodesic distances a node has with all the other nodes of the graph; the {\em Betweenness  Centrality} is calculated based on the number of shortest paths that pass through a node \cite{brandes2001faster}. However, as stated in \cite{borgatti2005centrality}, the importance of a node in a network cannot be determined without reference to how traffic or information flows through the network. For example, in a package delivery process, the essence of closeness is time-until-arrival, in contrast, the essence of betweenness is frequency of arrival. These interpretations do not seem to be able to represent the FRAM characteristics properly, where peripheral nodes may also result critical. \\ 
Figure \ref{fig:indegree} shows the graph obtained by encoding the FRAM specified in Table \ref{tab:functions} and Table \ref{tab:relations} available in the Appendix and ordering nodes in concentric range based on their DP value. In Tables \ref{tab:FanInDC} and \ref{tab:RelInDC}, node values are listed in decreasing order and we can observe that the functions exposing the highest values are  {\em F2:Coordinate service delivery}, {\em F16: Manage awareness and human behavior}, {\em F1: Delivery service},  {\em F24:Collect event information} and {\em F6: Coordinate emergency action}. \\

It is worth to notice that function $F16$, that is devoted to managing community behavior and awareness is considered one of the most critical in the network.  Such a result is not unexpected and confirms and formalizes what the stakeholder thoughts and the FRAM model are able to expresses only implicitly. Moreover, the DP centrality approach is able to represent the FRAM background functions\footnote{ FRAM Glossary - http://functionalresonance.com/a-fram-glossary.html} assigning 0 to the DP index.  \\
Regarding the relations affecting connected communities let us focus our attention on: {\em R106} (F13:User Behavior:F14:Input), {\em  R107} (F13:User Feedback:F14:Input),  {\em R108} (F14:User Behavior data:F16:Resources), {\em R109} (F14:User generated critical event detection:F2:Input), {\em  R110} (F14:User generated critical event detection:F6:Input), {\em R108} (F14:User generated service improvement suggestions:F24:Input); the analysis revels that R106, R107, R107 belong to the most important group, characterised by a DP equal to 0,64935; while the nodes R109 and R110 belongs to the second most important group, with DP equal to 0,58442, as reported in Table \ref{tab:RelInDC}.  \\

This analysis reveals that people-community related issues represent a critical aspect of the UTS resilience management.  In fact, if the variability of the outputs of those functions are exceeding the dampen capacity of the downstream functions,  such variability surplus is propagated in the system exhibiting a resonance behavior that can be preparatory for a disaster.
It is then clear that exploiting the IoE technologies enhances the capacity of such functions of damping the performance variability that can be generated by information delivery delay, misunderstandings, etc. 
The IoE adoption may speed up the co-production and the dissemination of information within the CCs (e.g. created during an emergency) and between these CCs and the first responders.\\

To reduce the size and complexity of the graph, one may also consider aggregating connections insisting on the same dimension, for instance, all the connections with the same origin function and destination function insisting on the same FRAM parameter. To manage size reduction consistently, standard approaches for multidimensional data such as OLAP Cubes \cite{ciferri2013cube} may be implemented, but ad hoc projection operators accounting data aggregation with a domain specific approach are also possible \cite{markines2009evaluating}.

\section{Quantify Functional Variability}
As previously discussed, in FRAM several dimensions with qualitative degrees are characterizing the FPV. However, even considering invariant the FPV of an origin function, the impact of this variability on the resonance vary based on the {\em dampening capacity} of the destination function. As stated in {\bf O2},  quantify such an impact is crucial. The current approaches do not offer any method to compare FPV over the capacity of the downstream functions of continuing to operate within normal variability in the face of varied inputs.  
Thus, we define the {\em function dampening capacity} (FDC) of a function $F$ as the capability of $F$, in a certain context, of absorbing the variability of the incoming input $I$ (changing conditions) maintaining  its output $O$ within acceptable/expected variability.  \\
We also argue that the factors composing the FDC index are the four properties considered for resilience assessment at system level (buffer capacity, flexibility, margin and tolerance) and introduced in \cite{Woltjer}. Hence, the FDC in a certain instant $t$ for a specific input $i$ is given by its function buffer capacities (FBC), function flexibility (FF), function margin (FM) and function tolerance (FT). 
However, in which degree those functions contribute to the FDC require further analysis and will be matter of next researches. \\

The formalisation we are proposing in this paper is aimed at quantifying the amount of FPV in upstream exceeding the FDC of a downstream function. In particular, we call this matching the {\em Variability  Rate} (VR). The VR expresses the amount of input variability dampened or absorbed by the downstream function avoiding effects on its subsequent outputs.

\subsection{Variability  Rate} \label{Sec:QFV}
A naif solution to quantitatively measure FDC is to interpret it as the inverse of a correlation. We compare the distribution of the performances of two connected functions along with a specific dimension. If we observe a correlation this can be considered a clue for an amplification effect on the downstream function, generated by the upstream function. Thus we have to consider that the downstream function has poor FDC. 
For example, in Table \ref{tab:perf} we list the performances of the {\em output} of functions F15, F2, and F6, measured on the timing dimension, using delay, expressed in hours, as value. Note that $F15$ is the upstream function of both $F2$ and $F6$. The Pearson correlation coefficient for $F15$ and $F2$ is $0.905$ while for $F15$  and $F6$ is $-0.153$. Thus we could conclude that $F6$ has a good FDC while $F2$ has not.

\begin{longtable}{p{.25\textwidth}  p{.03\textwidth} p{.03\textwidth} p{.03\textwidth} p{.03\textwidth} p{.03\textwidth} p{.03\textwidth} p{.03\textwidth} p{.03\textwidth} p{.03\textwidth} p{.03\textwidth}}
\toprule
$\mathbf{Function~Performances}$ & \multicolumn{10}{}{\mathbf{Delay in hours}} \\ \midrule
F15: Manage financial affaire  & 0 & 24 & 36 & 168 & 24 & 24 & 24 & 36 & 72 & 0 \\ \midrule
F2: Coordinate service delivery & 0 & 1  & 24 & 96  & 0  & 1  & 2  & 3  & 2  & 2 \\ \midrule
F6: Coordinate emergency action & 1 & 1  & 2  & 1   & 3  & 1  & 1  & 1  & 2  & 2 \\
\bottomrule    
\\
\caption{Performances of functions F15, F2, and F6 using delay in hours as value.}
\label{tab:perf}

\end{longtable}

This approach is, however, too much influenced by the internal variability of a function. As a matter of fact, our purpose is not accounting the variability in general but the variability generated by those performances that bring the function outside a margin of regular operation. This means we are not interested in accounting those performances that are within the margin.

The approach we are proposing is centred around the idea of computing how much a specific performance is differing form an expected value and comparing this value to the margin that delimits regular performances.  Formally this can be defined as in Equation \ref{eq:dev}, where $dev$ is the deviation, $x$ the observed performance, $e$ is the expected or more representative performance value and $m$ the margin of regular operation. When this fraction ranges in the interval $[-1, 1]$ the difference between the observed and the expected value is within the margin. Note that $e$ and $m$ could be defined as constant values or as the result of a function, for example, in a pawer low distribution, $m$ cloud be obtained by a function of $e$.

\begin{equation}\label{eq:dev}
dev = \Big| \frac{(x - e)}{m} \Big|
\end{equation}

If we use the {\em mean value} as $e$ and the {\em standard deviation} as $m$, our $dev$ is equivalent to the $z$-score. Clearly, the assumption of normal distribution required by the $z$-score is too restrictive for complex systems such as the UTS. In Table \ref{tab:perfdev} we computed the deviations of F15, F2 and F6 using two different approaches. In $dev^{a}$ we use the {\em median value} as $e$ and the {\em median absolute deviation}\footnote{The {\em median absolute deviation} of a series of observations is the median value of all the absolute deviations of each observation from the median value of the series. Formally this can be expressed as $mad = median(\bigcup_{i}^{n}|X_{i} - median(X)|)$.} as $m$, because the median is more robust than the mean to bias in skewed distributions. While in $dev^{b}$ we use ad-hoc thresholds: in particular, $0$ as $e$ and $24$ as $m$ for $F15$ or $1$ as $m$ for $F2$ and $F6$. For example, when the delay of $F2$ is $24$ hours, because the median value of the series of observations is $2$ and the median absolute deviation is $1$, $ dev^{a} = \frac{(24 - 2)}{1} = 22$. 

\begin{longtable}{p{.25\textwidth}  p{.03\textwidth} p{.03\textwidth} p{.03\textwidth} p{.03\textwidth} p{.03\textwidth} p{.03\textwidth} p{.03\textwidth} p{.03\textwidth} p{.03\textwidth} p{.03\textwidth}}
\toprule
$\mathbf{dev^{a}}$ & \multicolumn{10}{}{\mathbf{Delay in hours}} \\ \midrule
F15: Manage financial affaire  & 2 & 0 & 1 & 12 & 0 & 0 & 0 & 1 & 4 & 2 \\ \midrule
F2: Coordinate service delivery & 2 & 1  & 22 & 94  & 2  & 1  & 0  & 1  & 0  & 0 \\ \midrule
F6: Coordinate emergency action & 0 & 0  & 1  & 0 & 2  & 0  & 0  & 0  & 1  & 1 \\ \midrule
$\mathbf{dev^{b}}$ & \multicolumn{10}{}{\mathbf{Delay in hours}} \\ \midrule
F15: Manage financial affaire  & 0 & 1 & 1.5 & 7 & 1 & 1 & 1 & 1.5 & 3 & 0 \\ \midrule
F2: Coordinate service delivery & 0 & 1  & 24 & 96  & 0  & 1  & 2  & 3  & 2  & 2 \\ \midrule
F6: Coordinate emergency action & 1 & 1  & 2  & 1   & 3  & 1  & 1  & 1  & 2  & 2 \\
\bottomrule    
\\
\caption{Performances deviation for F15, F2, and F6, using $dev^{a}$ and $dev^{b}$.}
\label{tab:perfdev}

\end{longtable}

Using performance deviations we can now compute the FPV of an upstream function and the FDC of a downstream function, we can then quantify a matching between them to measure the VR.  

Since the variability of a function is exhibited in its {\em output} variability, the evaluation of the FDC of a downstream function can be performed by evaluating the variability of its output in relation to the variability of the input received. 


In particular, if the function has received inputs with a certain level of variability and the output of the function exhibits the same or increased level of variability, this means that the current FDC of the function was not enough to dampen incoming variability. The result is the variability propagation effect in the system that is called functional resonance in FRAM. Formally, we compute FPV as in Equation \ref{eq:fpv}, where $[1, ..., n]$ is the set of observations considered, i.e. distinct executions of the process. While FDC is given by summing the differences between deviations of the upstream and downstream functions for the same observation, referred as origin function ($O$) and destination function ($D$) in Equation \ref{eq:fdc}. The percentage of VR on a pair upstream, downstream function is then calculated as the ratio defined in Equation \ref{eq:vr}.

\begin{equation}\label{eq:fpv}
FPV(F) = \sum_{i=1}^{n} dev_{i}~|~dev_{i} \geq 1.
\end{equation}

\begin{equation}\label{eq:fdc}
FDC(D | O:q) = \sum_{i=1}^{n} dev_{O,i} - dev_{D,i}~|~dev_{O,i} \geq 1.
\end{equation}

\begin{equation}\label{eq:vr}
 \% VR_{O,D} = \frac{FDC(D|O)}{FPV(O)} \cdot 100.
 \end{equation}
 
 For example, as the $FPV_{dev^{a}}(F15)$ is $22$ and the $FDC_{dev^{a}}(F2 | F15)$ is $-76$, then the $VR_{dev^{a}}(F15, F2)$ we observe is  $-345\%$, indicating that the dumping capacity is negative, or, in other words, we have an amplification. On the contrary, the $FDC_{dev^{a}}(F6 | F15)$ is $18$, then the $VR_{dev^{a}}(F15, F6)$ we observe is  $81\%$, indicating that $F6$ can operate regularly even in presence of delays on $F15$. Note that these results are very much influenced by the margin of regular operation chosen, the $e$ and $m$ values in Equation \ref{eq:dev}. For example, we have $VR_{dev^{b}}(F15, F2) = -658\%$ and  $VR_{dev^{b}}(F15, F6) = 41\%$. However, it is also clear that the same orientation, in terms of positive or negative dumping capacity, is returned with different specification of Equation \ref{eq:dev}.

In order to visually represent the $VR$ as a ratio between $FDC$ and $FPV$, we propose to exploit an adjacency matrix, i.e. a square matrix such that its element $m_{i,j}$  is representing a relationship among two objects of the same class $P$; having the same set of objects encoded in both the rows and the columns of the matrix. This implies that the relationships encoded in an element $m_{i,j}$ are directed from the object $p_{i}$ to the object $p_{j}$, while the inverse relationship, directed from $p_{j}$ to $p_{i}$, is encoded in the element $m_{j,i}$. 


Let us, for example, encode  in a square matrix the relationships in $O \cup D$, i.e., the set of origin and destination functions. Using a Chord Diagram \cite{holten2006hierarchical}, we can display the relationships encoded in the matrix drawing arcs connecting the objects in  $O \cup D$ arranged radially around a circle. This approach is particularly appreciated by the end user as the radial order allows to estimate the global inter-connection density of the system as well as to explore each single relationship within a compact space.

\begin{figure}
    \centering
    \includegraphics[type=png,ext=.png,read=.png,width=12.5cm]
{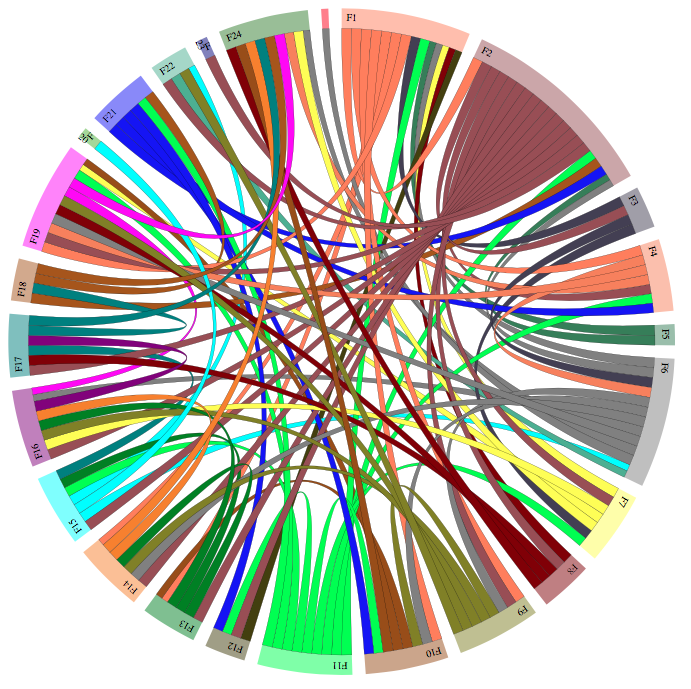}
    \caption{A Chord Diagram illustrating the VR characterising each relationships in $O \cup D$.}
    \label{fig:Chord}
\end{figure}

\begin{figure}
    \centering
    \includegraphics[type=png,ext=.png,read=.png,width=12.5cm]
{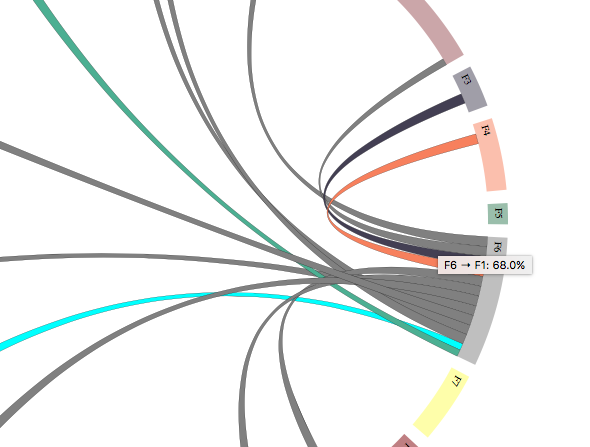}
    \caption{Interacting with the Chord Diagram to observe the VR characterising a specific relationship.}
    \label{fig:ChordZoom}
\end{figure}

Figure \ref{fig:Chord} illustrates the Chord Diagram that describes the FRAM specified in Table \ref{tab:functions} and Table \ref{tab:relations}. Note that this diagram allows interactive selection of relationships, to visualize their VR ratio, as illustrated in Figure \ref{fig:ChordZoom}.\\
%

\subsection{Encoding FRAM Variability Observations with fuzzy logic}

It is worth to notice that the FPV of a function is the result of multiple observations, possibly generated from different sources, where different encoding procedures may be applied. In many situations, it is naive to believe deriving the VR based on pure quantitative measurements. Hence, we propose to integrate the method described in Section \ref{Sec:QFV} with subjective reviews, using perception based definitions, as requested by {\bf O3}. The idea is that: when a quantitative value is not available, this value may be determined through the collection of a set of perception based valuations from domain experts. The imprecise nature of this kind of information is mitigated by focusing on the most representative opinion.

Many models are available for representing uncertain knowledge \cite{damiani2009toward}. In order to clarify the notion of uncertainty, it is important to distinguish between degrees of truth and degrees of uncertainty in the information. A degree of truth can be defined as the degree of compatibility between a statement and a fact. The uncertainty of a statement arises when there is no sufficient information to decide if a statement is true or false. In our study, the problem we want to resolve is related to the identification of a representative majority, the strength of this majority and the valuation reflecting the judgment of the majority \cite{ceravolo2007bottom}. The typical approach is to compute a value which synthesizes the opinion of the experts involved in the evaluation, but because different majorities are possible, we prefer to consider all of them, weighting their relevance based on how close they are to our idea of a full majority. In other words, we need to express degrees of truth. A formal method to deal with it is offered by Fuzzy Logic \cite{klir1995fuzzy}, which has been largely exploited for aggregating opinions \cite{herrera2014review}, votes \cite{ceravolo2005adding} or imprecise information \cite{herrera1997aggregation}, in general. 
In classical set theory, the characteristic function of a set $E \subseteq D$ is a function assigning $1$ or $0$ to each element of the domain $D$, depending on whether the element is in the subset $E$ or not. In Fuzzy Logic, the characteristic function of a set $\mu E$ returns for each element in the domain $D$ a value $\omega$ in the interval $[0, 1]$, which represents the grade of membership of this element in $E$.
Fuzzy sets can be used to partition the distribution of a variable in ranges corresponding to perception-based quantifiers, for instance {\em Very low}, {\em Low}, {\em Average}, {\em High}, {\em Very High}. These concepts, usually referred as linguistic variables, are exploited during reasoning to transform perception based observations into membership degrees.

To provide a formal framework to address this problem we adapted the approach followed in \cite{pasi2003modeling}. The general idea is to interpret the representative majority no longer as a single value, but as a fuzzy set. This fuzzy set includes all the possible subsets representatives of a majority within the collection of values expressing all the valuations. This require for the identification of both the strength of a majority and the synthesized value expressed by this majority. 

\begin{figure}[!ht]
    \centering
    \includegraphics[type=png,ext=.png,read=.png,width=13.5cm]
{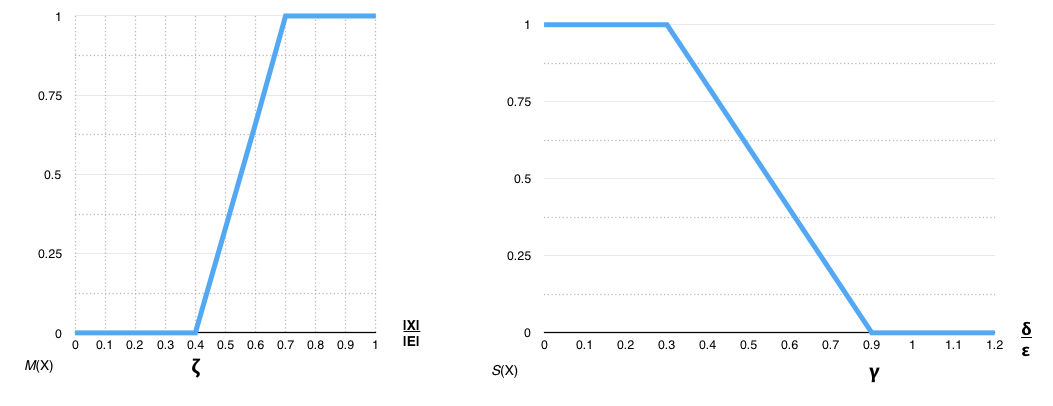}
    \caption{The characteristic functions $\mathcal{M}(X)$ and $\mathcal{S}(X)$.}
    \label{fig:majop}
\end{figure}

Once multiple observations from experts are gathers in a bag of valuations $E = \{ e_{1}, ..., e_{n} \}$, we are requested to define a characteristic function for identifying similar values. For example, asking that two values $e_{i}$ and $e_{j}$ are similar if their difference $\delta$ is not too far from a point of realisation $\epsilon$. Let us call $\mathcal{S}$ the characteristic function implementing this idea. The input of this function is $\frac{\delta}{\epsilon}$, this value is accounted as the intensity of the similarity,  in the interval $[0, 1]$, except for values $> \gamma$, an upper bound, that are accounted as $0$, as illustrated in Figure \ref{fig:majop}. Note that $\mathcal{S}$ is not a proper similarity as it is symmetric but not transitive, nevertheless this approach offers a very simple test condition. Another requirement is we define a characteristic function  $\mathcal{M}$ for identifying a majority. This function evaluates the cardinality of a subset $X_{i}  \subseteq E$ to define how intensively it can be considered a majority. $\mathcal{M}$ has a lower bound $\zeta$ that defines subsets we cannot consider a majority, moreover, it defines the intensity of a majority by computing $\frac{|X|}{|E|}$, as illustrated in Figure \ref{fig:majop}. Now we can identify a majority if $X_{i}$ contains elements that are similar and its cardinality satisfies our idea of being a majority. Let us formalise this notion by stating that a subset $X_{i}  \subseteq E$ is a majority with degree defined by a function $Maj(X_{i})$, where:

\begin{equation}
Maj(X_{i}) = min( \mathcal{M}(X_{i}), \mathcal{S}(X_{i})), with~\mathcal{S}(X_{i}) = Min_{e_{i}, e_{j} \in X_{i} } [\mathcal{S}(e_{i}, e_{j}) ].
\end{equation}

To consider all subsets of $E$ we have to generate the power set of $E$, i.e. we have to consider $2^{|E|}$ subsets. The intrinsic complexity of this approach may seem excessive, however note that the characteristic functions $\mathcal{S}$ and $\mathcal{M}$ act as filters for several subsets that are discarded.  Let us, for example, consider the following bag of opinions where values are drawn from a scale from 0 to 10:

\begin{equation}
E = \{1, 4, 4, 5, 6\}.
\end{equation}

We have $2^{5} = 32$ subsets. Nevertheless, taking as $\mathcal{M}$ the characteristic function illustrated in Figure \ref{fig:majop} all subsets with two elements are discarded because $ \frac{|X|}{|E|} < 0.4$. Moreover, shaping $\mathcal{S}$ as illustrated in Figure \ref{fig:majop}, any subset having any of its elements with a difference $ \geq 3$ is also discarded\footnote{When the distance equals to $3$ or more, $\frac{\delta}{\epsilon}$ is over $\gamma$.}. 

Thus the following are the only subsets for which $Maj(X_{i}) \neq 0: X_{1} = \{4, 4, 5\},  X_{2} =\{4, 4, 6\},  X_{3} =\{4, 5, 6\},  X_{4} =\{4, 5, 6\},  X_{5} =\{4, 4, 5, 6\}$. 
As illustrated in Table \ref{tab:majopex}, we can now compute $Majop(X_{i})$, i.e. the intensity of a majority, for each of them. The next step is computing the opinion expressed by each majority that we define as $Op(X_{i})$ = $AVG_{j}(x_{j} \in X_{i})$, i.e. the value averaging the opinions expressed in a majority. Based on $Maj(X_{i})$ we can weights $Op(X_{i})$. We define the weight of each majority as $W(X_{i})$, which is computed as:

\begin{equation}
W(X_{i}) = \frac{Maj(X_{i})}{\sum_{X_{j} \subseteq E} Maj(X_{j})}.
\end{equation}

Where $\sum_{X_{i}} W(X_{i}) = 1$. Now, the most representative value for $E$ can be computed as a weighted mean of the values expressed by each majority in $X_{i}  \subseteq E$. For example, 

\begin{equation}
Majop(E) = \sum_{X_{i}} W(X_{i}) \times Op(X_{i})
\end{equation}

Then, following our example, the proposed methodology determines the most representative value in $E$ as $4.75$, formally this is written as: $MajOp(E) = 4.75$. 
The method we presented can be applied to any value that is relevant for evaluating the FRAM. In Section \ref{sec:CS} we are using this method to evaluate the Z-score characterising the FDC of the relationships analysed in the scenario we investigated.

\begin{longtable}{p{.15\textwidth}  p{.15\textwidth} p{.15\textwidth} p{.15\textwidth} p{.15\textwidth} p{.15\textwidth} }
\toprule
$\mathbf{ X_{i}}$ & $\mathbf{\mathcal{S}(X_{i})}$ & $\mathbf{\mathcal{M}(X_{i})}$  & $\mathbf{Maj(X_{i})}$  & $\mathbf{Op(X_{i})}$ &   $\mathbf{W(X_{i})}$        \\ \midrule
$X_{1}$     & 0.99 & 0.66 & 0.66      & 4.33 &   0.22                \\ \midrule
$X_{2}$     & 0.66 & 0.66 & 0.66      & 4.66 &   0.22              \\ \midrule
$X_{3}$     & 0.66 & 0.66 & 0.66      & 5 &        0.22            \\ \midrule
$X_{4}$     & 0.66 & 0.66 & 0.66      & 5 &        0.22            \\ \midrule
$X_{5}$     & 0.66 & 1 & 0.66          & 4.75 &    0.22            \\ \midrule
$MajOp(E)$     & 4.75 &  &  & &              \\ 
\bottomrule    
\\
\caption{Main steps required to compute $MajOp(E)$ }
\label{tab:majopex}

\end{longtable}

\section{Case Study}\label{sec:CS}

In emergency  it is very important to manage CC behaviors effectively in order to reduce the impact generated also by wrong decisions. To this end the UTS needs to properly address  the community related system functions as: {\em F16: Manage awareness and user behavior} . 

In this scenario, it is necessary to support community members in taking right decisions for their safety while addressing their mobility goals. The methodology considered in RESOLUTE is inspired by Bungay's Directed Opportunism approach \cite{bungay} since it represents the main shift of the strategy, from centralised ``command and Control" to ``Mission Control"  and it can be  considered more appropriate  to support  the community self-resilience. The Bungay's approach is a control loop composed by Outcome (e.g. reduction of the car presence  in the affected area through redirection of the traffic flow towards other part of the road network), Plans (e.g. application of rerouting strategy), Actions (e.g. sets of traffic lights cycle, close streets, send recommendation message to city panels) steps. The objective of the approach is to left people free to take opportunistic decisions exploiting their local knowledge that is considered more accurate respect to the centralised one. However, to guarantee that such decisions are actually appropriate to the actual condition, the following gaps need to be crossed:

\begin{enumerate}
\item \textbf{Knowledge Gap}: the delta between what we would like to know and what we actually know.
\item \textbf{Alignment Gap}: the difference between what we want people to do and what they actually do.
\item \textbf{Effects Gap}: the difference between the expected and the actual results of our actions.
\end{enumerate}

When these gaps are encountered, the intuitive response is to seek increasing the control on these areas by gathering more detailed information, providing more detailed instruction, and installing tighter controls. These intuitive responses typically have the opposite effect of their intent, creating greater confusion and entropy. In fact, in order to reduce these gaps it is necessary to apply the following actions:

\begin{enumerate}
\item Do not command more than is necessary or plan beyond the circumstances you can foresee (Knowledge gap).
\item Communicate to every one as much of the higher intent as is necessary to achieve the purpose (Alignment gap).
\item Being sure everyone is empowered to make decisions within bounds (Effects gap). 
\end{enumerate}

The data used for the analysis are directly derived by the 1st RESOLUTE stakeholder workshops held in Florence in December 2015, where several scenarios were analyzed (water bomb/flash flooding, large yard, river flooding, car accidents, etc.).  For use case addressed  in the present work, we take into account the flash flooding extreme event. 
This kind of event is charactersied by sudden, unpredictable and localised (radius of few kilometers) extreme heavy rain that is capable to badly affect UTS operations (cars speed reduction/ blocked, car accidents for reduced visibility or roads grip, fall trees, traffic lights out of order, underpasses flooded, etc.) in a specific part of a city.

This scenario is assessed considering cases in which: a) an UTS where the IoE and Connected Community are integrated in the system, and b) an UTS where such facilities are not exploited. In particular, as we explained, the IoE and Connected Community emerging concepts consider the people part of the system (e.g., Smart City). In this system, the people can be considered as always connected through their personal devices to informal, fluid and/or thematic communities exploiting different communication infrastructures (e.g. city free wifi or 4G/LTE).   

This assumption is justified by the existence of WAN communication infrastructures with several overlaps (e.g., LoRaWAN) that are able to guarantee connectivity also in case of disaster. On the other hand, it is evident that in case of extreme disruption where all the communication and electrical infrastructures are damaged, the IoE and CC facilities cannot be exploited until such infrastructures have been restored.

The people want to access tailored and context aware data and services and stay in contact with the other community members every time everywhere to exchange information, suggestions, to share their opportunistic decisions, etc.. During the emergency such connections can be exploited to alert first responders about the position of the disaster, or can be used by the authorities to send personalized messages according to the 4R approach (right person at right time, in the right place, through the right channel) or generic messages for the entire community to speed up the dissemination of critical information. 
In the following tables,  two scenarios mapped to the UTS FRAM function relations affected, are reported.

\begin{longtable}{p{.18\textwidth}  p{.30\textwidth}  p{.30\textwidth} p{.10\textwidth}}
\toprule

Relations & Standard Scenario & CC Scenario & Gap    \\ \midrule

{\em F13:User Behavior:F14:input}
& User behaviors signals are basically captured by security/surveillance video systems with limits in view extensions, mass quantification, people velocity and direction, information processing, sharing capability, etc. 
During a water bomb people tends to reduce the speed up to block the local variability. Such event is very difficult to be propagated to the other drivers that are reaching the same area, thus they will be surprised by the event even if they arrive minutes after. In order to escape from the traffic jam, people tends to broke roads rules with U turns, reverse gear, double parking, etc. People tends to apply heuristics combined with local knowledge and risk perception to continue of addressing their movement objectives (e.g., go back home from work).
& User as a sensor concept allow real-time people movement tracking through 3/4G, Public Wifi, bluetooth, etc. Positions, velocities, trajectories  can be calculated and predicted in close to real-time. Such information can be shared with multi decision makers through dashboards. Providing personalised real-time and contextualised  information and suggestions, allows people to take a decisions in due time (e.g. avoiding traffic jam if the driver is arriving in the affected area) \cite{fgcs}.    
&  Knowledge, Effects Gap\\ \midrule

{\em F13:User feedback:F14:input}
& People communicate with authorities or UTS operators through call centres, email or directly with the transport employees (e.g. bus drivers). Such channels are basically used for caring.

& People can post pictures and videos of a critical events on social networks in real-time with a impressive dissemination capability. Applications like WhatsApp, Telegram or XMPP allow instant multimedia messaging with groups and communities.  
& Effect \\ \midrule
 
 {\em F14:User behavior data:F16:input}
& Data about people behavior and movement are usually derived from slow dynamic data like seasonality of touristic visits,  daily traffic pressure for work entrance/exit,  etc. Such data are  indicators that are useful for risk assessment while for real-time emergency decisions (e.g. directing first responders in a specific area)  more dynamic and fresh information are necessary.
& IoE and CC allow the possibility to deliver maps of people behaviors integrating different channels from social networks (georeferenced Tweets of pictures in Flickr), GPS signals from always connected smart devices, etc. 
& Knowledge, Alignment \\ \midrule
        
{\em F14:User generated critical event detection:F6:input}
{\em F14:User generated critical event detection:F2:input}
& During the emergency people alert first responders about a critical event basically through telephone call, but misunderstandings, partial descriptions of the scenario, wrong perception given by stressful situation, language gaps, etc. impact on the precision of the description and thus on the effectiveness of the emergency action.  Moreover, in Italy there are several emergency numbers (113- Police, 118 Ambulance, 115 Fire brigades, etc.) that people can call for  an emergency. This fragmentation determine multiple calls for the same event increasing the noise.
&  The possibility of taking and sharing pictures, videos, voice and text messages in real-time improve the quality of the information and the event understanding.
& Knowledge \\ \midrule

{\em F14:User generated service improvement suggestions:F24}
& Improvements are basically driven by post-event accident analysis and it is carried out by experts in the field without any participatory approach.  
& Opening a direct digital channel towards the users/citizens to collect their suggestions improve the understandability of the system usage and perception. Sentiment, clustering and statistical analysis on such a database can extract unexpected knowledge.

& Knowledge\\ \midrule

{\em  F16:User generated service improvement suggestions:F13}
& Authorities and UTS have a very partial idea in which place and what the people are doing when a critical event happen. Information is spread in broadcast using massive and pre registered telephone calls (if the people is registered to alerting service), megaphone, Tv and radio channels, variable message panels, etc.. The messages are usually synthetic with very few  information and reach the intent is to reach more people as possible without any filter. Generally, people that are not prepared to cope with emergency, thus they tend to react according to their heuristics, emotion, past experience, local knowledge, etc. with an high level of uncertainty (e.g. people that try to save the car in the box during the flooding).
& The capability of exploiting smart devices as a personal end point to communicate  the right message to the right person at the right time through the right channel - The 4R approach - represents a relevant  improvement offered by IoE to enhance UTS resilience addressing the human-social side. 
&  \\ \bottomrule    
\\
\caption{Relationships under investigation in the RESOLUTE case study.}
\label{tab:FanPC}
\end{longtable}

\subsection{Quantitative development of the case study}

To develop a quantitative evaluation about the impact of migrating a system into a CC scenario we applied the methodology described in Section 5.2.\\ 



The first step foresees the collection of the VR assessment provided by the experts. 

In fact,  the FRAM is basically a qualitative method, the VR can be evaluated looking at the expert judgment in {\em input} about the variability of function $F$ and the expert judgment about the variability of its subsequent {\em output}. Since, such linguistic variables express a variability range, the estimation of VR provided by the experts for each relations, can vary from one function to another even if the input and output variability judgments are similar.
For instance, if the timing variability in {\em input}  of a function F is evaluated as ``too late" and the variability of $F$  {\em output} is evaluated as ``in time", this does not mean that the VR estimated by the expert for $F$, should be equal to another function $G$ that has the same input and output variability judgments. In fact, VR  estimation of a function $F$ should take into account its function buffer capacities  (FBC), function flexibility (FF), function margin (FM) and function tolerance (FT). This means that, even if  $F${\em output} is judged ``in time", the VR estimation is influenced by the assessment of the function aspects status at the instant $t$  considered.
The VR estimation provided by 8 local experts (that has been selected carefully of the bases of their experience and decision level), against the flash flooding analysis in Florence is a number from 0 to 10 representing the VR percentage (for instance 1= 10\%). The $MajOp(E)$ value is calculated based on the membership functions illustrated in Figure \ref{fig:majop}.\\

{\em F13:User Behavior:F14:input}\\
\noindent\rule{12cm}{0.4pt}\\
{\em Standard Scenario}: $E = \{ 1, 0, 2, 2, 4, 5, 2, 2 \}$;   $MajOp(E)=1.15$; $VR=11.5\% $\\
{\em CC Scenario}: $E_{CC} = \{ 9, 8, 9, 7, 9, 8, 7, 7  \}$; $MajOp(E_{CC})=8$; $VR=80\%$\\

{\em F13:User feedback):F14:input}\\ 
\noindent\rule{12cm}{0.4pt}\\
{\em Standard Scenario}: 
$E = \{ 1, 0, 2, 2, 1, 0, 4, 2 \}$; $MajOp(E)=1.15$; $VR=11.5\%$\\
 {\em CC Scenario}: 
$E_{CC} = \{ 9, 6, 6, 9, 8, 5,6, 7  \}$; $MajOp(E_{CC})=6.4$; $VR=64\%$\\
 
 {\em F14:User behavior data:F16:resources}\\ 
 \noindent\rule{12cm}{0.4pt}\\
{\em Standard Scenario}:
 $E = \{ 1, 0, 0, 2, 0, 0, 2, 2 \}$; $MajOp(E)=0.87$; $VR=8.7\%$\\ 
 {\em CC Scenario}:
$E_{CC} = \{ 10, 10, 10, 9, 10, 9, 9, 10  \}$; $MajOp(E_{CC})=9.62$; $VR=96.2\%$\\ 
        
{\em F14:User generated critical event detection:F6:input}\\ 
{\em F14:User generated critical event detection):F2:input}\\ 
\noindent\rule{12cm}{0.4pt}\\
{\em Standard Scenario}: 
$E = \{ 5, 4, 6, 2, 4, 5, 3, 6 \}$; $MajOp(E)=4.82$; $VR=48.2\%$\\ 
{\em CC Scenario}:
$E_{CC} = \{ 10, 10, 9, 9, 8, 10, 10, 6 \}$; $MajOp(E_{CC})=9$; $VR=90\%$\\ 

{\em F14:User generated service improvement suggestions:F24:input}\\ 
\noindent\rule{12cm}{0.4pt}\\
{\em Standard Scenario}:
$E = \{ 2, 0, 2, 0, 1, 2, 1, 2 \}$; $MajOp(E)=1.25$; $VR=12.5\%$\\ 
{\em CC Scenario}:
$E_{CC} = \{ 6, 6, 7, 8, 6, 5, 7, 7 \}$; $MajOp(E_{CC})=6.5$; $VR=65\%$ \\

{\em F16:Warnings -Alerts:F13:resources}\\ 
\noindent\rule{12cm}{0.4pt}\\
{\em Standard Scenario}:
$E = \{ 1, 0, 2, 2, 4, 5, 2, 2 \}$; $MajOp(E)=1.65$; $VR=16.5\%$\\ 
{\em CC Scenario}:
$E_{CC}= \{ 10, 10, 10, 9, 10, 10, 9, 10 \}$; $MajOp(E_{CC})=9.75$; $VR=97.5\%$\\ 

{\em  F16:Advice - Recommendation Alert:F13:resources}\\ 
\noindent\rule{12cm}{0.4pt}\\
{\em Standard Scenario}:
$E = \{ 1, 1, 3, 0, 1, 2, 0, 1 \}$; $MajOp(E)=0.99$; $VR=9.9\%$ \\ 
{\em CC Scenario}:
$E_{CC} = \{ 10, 10, 10, 9, 10, 9, 9, 10 \}$; $MajOp(E_{CC})= 9.62$; $VR=96.2\%$\\ 

The results show that, according the expert judgments, the percentage of the variability that can be potentially absorbed by the functions with the CC and IoE technologies, is significantly higher respect to the same functions that operate without such facilities (standard scenario). This means that thanks to the introduction of such new technologies the capacity to manage people and community during the emergency is inherently enhanced and  the propagation of variability in the system is prevented or mitigated. 
In fact,  VR is influenced by the FDC as well as the Output variability distribution.  The VR score may result high also when the Output variability of the upstream is significantly reduced given the FDC of the downstream function. This could happen for instance when the communication processes managed by F16 becomes pervasive, ubiquitous and  personalized thanks to the IoE and CC technologies.  Thus the contribution of the CC and IoE is twofold: a) on FDC enhancement and b) on Output variability reduction. 
In this perspective, the evidence shows that introducing IoE and CC to enhance resilience in UTS  represents an option whose benefits value from 6 to 8 times the VR increment. 

\section{Conclusions}
Connected Communities and the related enabling technologies (Personal Smart Devices, multiple communication networks as WiFi, Bluetooth, LTE, Smart Sensors, etc.) set the scene of a new class of emergency and decision support systems based on knowledge, real-time situational awareness and personalised communication. 
In the present article, the Connected Community concept has been applied to the UTS resilience scenario to demonstrate the capability of such a concept in addressing the human-social side of the emergency in a more effective way, enhancing the resilience of the system as a whole.
To this end, we started from the RESOLUTE Resilience Management Guidelines, where a Critical Infrastructure reference model based on FRAM has been proposed.  Then we have developed a new method to analyse and quantify function's variability as a method to move towards resilience quantification. The application of a network science approach to the FRAM model,  has revealed what are the most critical functions in the system, while a method based on deviation score, was used to define the general principle for variability quantification. Since in FRAM the assessment is based on qualitative judgment, a  fuzzy logic based method was proposed to translate perception based observations into a quantification of the VR. In particular, a fuzzy notion of majority was adopted to guarantee a representative value. 
A scenario from those explored in RESOLUTE project, the water bomb, offered us the opportunity to compare and quantify the variability of those functions devoted to manage community aspects in UTS, considering two different contexts: where CC and IoE are deployed and where they are not (standard situation).\\
The outcomes obtained from the expert judgments on VR estimation reveal a remarkable differences between the two cases. This result shows that a technological upgrade of the UTS community-related functions towards the IoE and CC, would have an impact on the system resilience as a whole. In fact, such a VR enhancement in a specific connection between two functions, may act as an adaptive levee through the reduction of output variability of the upstream function or the enhancement of the damping capacity of the downstream function. Thus, the propagation of the variability in the system through function interdependencies that may trigger the resonance effect is prevented or mitigated within a threshold of acceptance.

We can conclude that the introduction of IoE and CC in UTS domain allows for the implementation of the next generation decision support systems, able to gather any kind of data generated by smart cities. 
The possibility of knowing where people are situated in a specific moment, their direction, velocity, and concentration, as well as, the possibility to reach them collectively or personally, in every time, everywhere, with tailored information, enhances the effectiveness of respond and recovery actions during emergencies. Nevertheless, the assessment of the global properties of a system, such as resilience, asks for the consistent integration of quantitative and perception based evaluations. Future researches will focus on these aspects by investigating alternative formalizations of the FDC, the quantification of the effects of mechanisms to simulate the variability propagation within the system, the development of a decision support system able to predict and provide recommendations on optimal resource allocation and technology upgrade to enhance VR in critical interdependencies.  


\section*{Acknowledgement}
This work has been supported by the RESOLUTE project (www.RESOLUTE-eu.org) and has been funded within the European Commission H2020 Programme under contract number 653460. This paper expresses the opinions of the authors and not necessarily those of the European Commission. The European Commission is not liable for any use that may be made of the information contained in this paper.

\addcontentsline{toc}{chapter}{References}
\newpage
\bibliography{biblio10.3}
\bibliographystyle{ACM-Reference-Format-Journals}                             
                             
\appendix
\section{Appendices}
\subsection{The RESOLUTE FRAM Model}

\begin{longtable}{p{.15\textwidth}  p{.55\textwidth} }
\toprule
ID & Function                                   \\ \midrule
F1              & Deliver service                           \\ \midrule
F2              & Coordinate service delivery                \\ \midrule
F3              & Manage human resources                     \\ \midrule
F4              & Training staff                             \\ \midrule
F5              & Supply resources                           \\ \midrule
F6              & Coordinate emergency action                \\ \midrule
F7              & Repair/restore operations                   \\ \midrule
F8              & Maintain physical/cyber infrastructure    \\ \midrule
F9              & Manage ICT resources                       \\ \midrule
F10             & Monitor safety and security                \\ \midrule
F11             & Regulate domain and operation              \\ \midrule
F12             & Define procedures                          \\ \midrule
F13             & Use of the service                         \\ \midrule
F14             & Monitor user generated feedback            \\ \midrule
F15             & Manage financial affaire                   \\ \midrule
F16             & Manage awareness and user behaviour        \\ \midrule
F17             & Develop strategic plan                     \\ \midrule
F18             & Provide adaptation and improvement insight \\ \midrule
F19             & Monitor operation                          \\ \midrule
F20             & Supply financial resources                 \\ \midrule
F21             & Perform risk assessment                    \\ \midrule
F22             & Monitor resource availability              \\ \midrule
F23             & Provide risk warning                       \\ \midrule
F24             & Collet event information                   \\ \midrule
F25             & Fight the emergency                   \\ 
\bottomrule    
\\
\caption{List of functions included in the FRAM}
\label{tab:functions}

\end{longtable}

\begin{longtable}{p{.06\textwidth} p{.13\textwidth} p{.40\textwidth} p{.13\textwidth} p{.15\textwidth}} 
\toprule

ID 	&		 Origin Function 	&	 Qualified Name                                                    	&		 Destination Function 	&	 Aspect       	\\ \midrule
R26            	&	F1	&	 Infrastructure performance                          	&	F	14	&	 Input        	\\ \midrule
R27            	&	F1	&	 Service                                             		&	F13	&	 Resources    	\\ \midrule
R28            	&	F1	&	 Service performance                                 	&	F19	&	 Input        	\\ \midrule
R29            	&	F1	&	 Service\_Safety\_Security\_performance              	&	F	10	&	 Input        	\\ \midrule
R30            	&	F2	&	 Operation HR plan                                   	&	F3	&	 Input        	\\ \midrule
R31            	&	F2	&	 Operation plan                                      	&	F16	&	 Input        	\\ \midrule
R32            	&	F2	&	 Operation plan                                      	&	F16	&	 Control      	\\ \midrule
R33            	&	F2	&	 Operation plan                                      	&	F12	&	 Input        	\\ \midrule
R34            	&	F2	&	 Operation plan                                      	&	F22	&	 Resources    	\\ \midrule
R35            	&	F2	&	 Operation Restore service request                   	&	F7	&	 Input        	\\ \midrule
R36            	&	F2	&	 Service delivery plan                               	&	F1	&	 Input        	\\ \midrule
R37            	&	F2	&	 Service delivery plan                               	&	F13	&	 Resources    	\\ \midrule
R38            	&	F2	&	 Service delivery plan                               	&	F16	&	 Resources    	\\ \midrule
R39            	&	F2	&	 Service improvement plan                            	&	F8	&	 Input        	\\ \midrule
R40            	&	F2	&	 Training staff requirements                         	&	F4	&	 Input        	\\ \midrule
R41            	&	F3	&	 human resources availability                        	&	F1	&	 Resources    	\\ \midrule
R42            	&	F3	&	 human resources availability                        	&	F6	&	 Resources    	\\ \midrule
R43            	&	F3	&	 human resources availability                        	&	F7	&	 Resources    	\\ \midrule
R44            	&	F4	&	 Staff trained                                       	&	F1	&	 Resources    	\\ \midrule
R45            	&	F4	&	 Staff trained                                       	&	F19	&	 Resources    	\\ \midrule
R46            	&	F4	&	 Staff trained                                       	&	F19	&	 Precondition 	\\ \midrule
R47            	&	F4	&	 Staff trained                                       	&	F6	&	 Precondition 	\\ \midrule
R48            	&	F4	&	 Training performance data                           	&	F24	&	 Input        	\\ \midrule
R49            	&	F5	&	 Supply resources                                    	&	F1	&	 Resources    	\\ \midrule
R50            	&	F5	&	 Supply status                                       	&	F2	&	 Resources    	\\ \midrule
R51            	&	F6	&	 Emergency HR request                                	&	F3	&	 Input        	\\ \midrule
R52            	&	F6	&	 Emergency response command                          	&	F25	&	 Input        	\\ \midrule
R53            	&	F6	&	 Emergency response data                             	&	F10	&	 Resources    	\\ \midrule
R54            	&	F6	&	 Emergency response data                             	&	F24	&	 Input        	\\ \midrule
R55            	&	F6	&	 Emergency response plan                             	&	F2	&	 Resources    	\\ \midrule
R56            	&	F6	&	 Emergency response plan                             	&	F10	&	 Input        	\\ \midrule
R57            	&	F6	&	 Emergency response status                           	&	F2	&	 Resources    	\\ \midrule
R58            	&	F6	&	 Emergency response status                           	&	F16	&	 Input        	\\ \midrule
R59            	&	F6	&	 Emergency response status                           	&	F10	&	 Resources    	\\ \midrule
R60            	&	F6	&	 Emergency response status                           	&	F19	&	 Resources    	\\ \midrule
R61            	&	F6	&	 Emergency response status                           	&	F1	&	 Resources    	\\ \midrule
R62            	&	F7	&	 Operation Restore service plan                      	&	F2	&	 Resources    	\\ \midrule
R63            	&	F7	&	 Operation Restore service plan                      	&	F16	&	 Input        	\\ \midrule
R64            	&	F7	&	 Operation restore/repair performance data           	&	F24	&	 Input        	\\ \midrule
R65            	&	F7	&	 Operation restore/repair status                     	&	F2	&	 Resources    	\\ \midrule
R66            	&	F7	&	 Operation restore/repair status                     	&	F16	&	 Input        	\\ \midrule
R67            	&	F7	&	 Operation restored/repaired                         	&	F1	&	 Precondition 	\\ \midrule
R68            	&	F7	&	 Operation restored/repaired                         	&	F2	&	 Input        	\\ \midrule
R69            	&	F7	&	 Operation restored/repaired                         	&	F19	&	 Precondition 	\\ \midrule
R70            	&	F8	&	 Infrastructure installed maintained                 	&	F1	&	 Precondition 	\\ \midrule
R71            	&	F8	&	 Infrastructure resotore/repair performance data     	&	F24	&	 Input        	\\ \midrule
R72            	&	F8	&	 Infrastructure resotore/repair plan                 	&	F2	&	 Resources    	\\ \midrule
R73            	&	F8	&	 Infrastructure restored repaired status             	&	F2	&	 Resources    	\\ \midrule
R74            	&	F8	&	 Infrastructure restored/repaired                    	&	F1	&	 Precondition 	\\ \midrule
R75            	&	F8	&	 Infrastructure restored/repaired                    	&	F2	&	 Input        	\\ \midrule
R76            	&	F8	&	 Infrastructure restored/repaired                    	&	F2	&	 Resources    	\\ \midrule
R77            	&	F9	&	 ICT infrastructures                                 	&	F1	&	 Precondition 	\\ \midrule
R78            	&	F9	&	 ICT infrastructures                                 	&	F2	&	 Resources    	\\ \midrule
R79            	&	F9	&	 ICT infrastructures                                 	&	F16	&	 Resources    	\\ \midrule
R80            	&	F9	&	 ICT infrastructures                                 	&	F10	&	 Resources    	\\ \midrule
R81            	&	F9	&	 ICT infrastructures                                 	&	F19	&	 Resources    	\\ \midrule
R82            	&	F9	&	 ICT infrastructures                                 	&	F22	&	 Resources    	\\ \midrule
R83            	&	F9	&	 ICT infrastructures                                 	&	F14	&	 Resources    	\\ \midrule
R84            	&	F9	&	 ICT infrastructures                                 	&	F6	&	 Resources    	\\ \midrule
R85            	&	F9	&	 ICT resource performance                            	&	F22	&	 Input        	\\ \midrule
R86            	&	F10	&	 Safety Security control                             	&	F1	&	 Control      	\\ \midrule
R87            	&	F10	&	 Safety Security control                             	&	F13	&	 Control      	\\ \midrule
R88            	&	F10	&	 Safety Security control                             	&	F19	&	 Control      	\\ \midrule
R89            	&	F10	&	 Safety Security critical event detection            	&	F6	&	 Input        	\\ \midrule
R90            	&	F10	&	 Safety Security performance data                    	&	F24	&	 Input        	\\ \midrule
R91            	&	F11	&	 Law                                                 	&	F1	&	 Control      	\\ \midrule
R92            	&	F11	&	 Law                                                 	&	F15	&	 Control      	\\ \midrule
R93            	&	F11	&	 Law                                                 	&	F2	&	 Control      	\\ \midrule
R94            	&	F11	&	 Law                                                 	&	F19	&	 Control      	\\ \midrule
R95            	&	F11	&	 Law                                                 	&	F7	&	 Control      	\\ \midrule
R96            	&	F11	&	 Safety regulation                                   	&	F21	&	 Resources    	\\ \midrule
R97            	&	F11	&	 Safety regulation                                   	&	F4	&	 Input        	\\ \midrule
R98            	&	F11	&	 Safety regulation                                   	&	F12	&	 Resources    	\\ \midrule
R99            	&	F11	&	 Safety regulation                                   	&	F10	&	 Control      	\\ \midrule
R100           	&	F11	&	 Standards                                           	&	F1	&	 Control      	\\ \midrule
R101           	&	F11	&	 Standards                                           	&	F2	&	 Control      	\\ \midrule
R102           	&	F11	&	 Standards                                           	&	F19	&	 Control      	\\ \midrule
R103           	&	F11	&	 Standards                                           	&	F7	&	 Control      	\\ \midrule
R104           	&	F12	&	 Procedure                                           	&	F1	&	 Control      	\\ \midrule
R105           	&	F13	&	 Revenues                                            	&	F15	&	 Resources    	\\ \midrule
R106           	&	F13	&	 User Behaviour                                      	&	F14	&	 Input        	\\ \midrule
R107           	&	F13	&	 User feedback                                       	&	F14	&	 Input        	\\ \midrule
R108           	&	F14	&	 User behaviour data                                 	&	F16	&	 Resources    	\\ \midrule
R109           	&	F14	&	 User generated critical event detection             	&	F2	&	 Input        	\\ \midrule
R110           	&	F14	&	 User generated critical event detection             	&	F6	&	 Input        	\\ \midrule
R111           	&	F14	&	 User generated service improvement suggestions      	&	F24	&	 Input        	\\ \midrule
R112           	&	F15	&	 Budget                                              	&	F2	&	 Resources    	\\ \midrule
R113           	&	F15	&	 SLA(Service Level Agreement)                        	&	F2	&	 Control      	\\ \midrule
R114           	&	F15	&	 SLA(Service Level Agreement)                        	&	F22	&	 Resources    	\\ \midrule
R115           	&	F15	&	 SLA(Service Level Agreement)                        	&	F6	&	 Control      	\\ \midrule
R116           	&	F16	&	 Early warnings                                      	&	F13	&	 Resources    	\\ \midrule
R117           	&	F16	&	 Service status                                      	&	F13	&	 Resources    	\\ \midrule
R118           	&	F17	&	 Develop strategic plan                              	&	F15	&	 Input        	\\ \midrule
R119           	&	F17	&	 Strategic plan                                      	&	F2	&	 Resources    	\\ \midrule
R120           	&	F17	&	 Strategic plan                                      	&	F16	&	 Input        	\\ \midrule
R121           	&	F17	&	 Strategic plan                                      	&	F8	&	 Input        	\\ \midrule
R122           	&	F17	&	 Strategic plan                                      	&	F18	&	 Input        	\\ \midrule
R123           	&	F17	&	 Strategic plan                                      	&	F24	&	 Control      	\\ \midrule
R124           	&	F18	&	 Event\_analysis\_insights                           	&	F21	&	 Input        	\\ \midrule
R125           	&	F18	&	 Knowledge base                                      	&	F24	&	 Resources    	\\ \midrule
R126           	&	F18	&	 Service sustained adaptability improvement insights 	&	F2	&	 Input        	\\ \midrule
R127           	&	F18	&	 System Sustained adaptability insights              	&	F17	&	 Input        	\\ \midrule
R128           	&	F19	&	 Install Maintenance requirement                     	&	F8	&	 Input        	\\ \midrule
R129           	&	F19	&	 Install Maintenance requirement                     	&	F2	&	 Input        	\\ \midrule
R130           	&	F19	&	 Operation Critical event detection                  	&	F16	&	 Input        	\\ \midrule
R131           	&	F19	&	 Operation Critical event detection                  	&	F6	&	 Input        	\\ \midrule
R132           	&	F19	&	 Operation performance monitoring data               	&	F16	&	 Resources    	\\ \midrule
R133           	&	F19	&	 Operation performance monitoring data               	&	F24	&	 Input        	\\ \midrule
R134           	&	F19	&	 Operation requirements                              	&	F2	&	 Input        	\\ \midrule
R135           	&	F20	&	 Funds                                               	&	F15	&	 Resources    	\\ \midrule
R136           	&	F21	&	 Risk assessment report                              	&	F4	&	 Input        	\\ \midrule
R137           	&	F21	&	 Risk assessment report                              	&	F2	&	 Resources    	\\ \midrule
R138           	&	F21	&	 Risk assessment report                              	&	F12	&	 Input        	\\ \midrule
R139           	&	F21	&	 Risk assessment report                              	&	F10	&	 Input        	\\ \midrule
R140           	&	F22	&	 Energy supply report                                	&	F2	&	 Input        	\\ \midrule
R141           	&	F22	&	 Resource supplied Critical event detection          	&	F2	&	 Input        	\\ \midrule
R142           	&	F22	&	 Resource supplied Critical event detection          	&	F6	&	 Input        	\\ \midrule
R143           	&	F23	&	 Official risk warning                               	&	F2	&	 Input        	\\ \midrule
R144           	&	F24	&	 Knowledge base 	&	F18	&	 Resources    	\\ \midrule
R145           	&	F2	&	 Operation Plan  	&	F9	&	 Input    	\\ \midrule
R146           	&	F5	&	 Supply Resources	&	F9	&	 Resources    	 \\

\bottomrule
\\
\caption{List of relations included in the FRAM}
\label{tab:relations}
\end{longtable}

\subsection{Analytics on the FRAM Model}

\begin{longtable}{p{.15\textwidth}  p{.55\textwidth} }
\toprule
Function ID & Degree Prestige Centrality   \\ \midrule
F2	&	10,13	\\ \midrule
F16	&	5,3896	\\ \midrule
F1	&	5,3247	\\ \midrule
F24	&	4,2208	\\ \midrule
F6	&	3,5714	\\ \midrule
F19	&	3,3117	\\ \midrule
F10	&	2,8571	\\ \midrule
F13	&	2,7922	\\ \midrule
F14	&	2,4675	\\ \midrule
F22	&	1,6883	\\ \midrule
F15	&	1,2338	\\ \midrule
F8	&	1,1039	\\ \midrule
F7	&	1,039	\\ \midrule
F3	&	0,97403	\\ \midrule
F12	&	0,84416	\\ \midrule
F4	&	0,77922	\\ \midrule
F21	&	0,71429	\\ \midrule
F9	&	0,64935	\\ \midrule
F25	&	0,64935	\\ \midrule
F18	&	0,58442	\\ \midrule
F17	&	0,45455	\\ \midrule
F5	&	0	\\ \midrule
F11	&	0	\\ \midrule
F20	&	0	\\ \midrule
F23	&	0	\\ 

\bottomrule    
\\
\caption{Functions ordered by Degree Prestige Centrality}
\label{tab:FanInDC}

\end{longtable}

\begin{longtable}{p{.20\textwidth}  p{.50\textwidth} }
\toprule
Relationship ID & Degree Prestige Centrality   \\ \midrule
R26	&	0,64935	\\ \midrule
R27	&	0,64935	\\ \midrule
R28	&	0,64935	\\ \midrule
R29	&	0,64935	\\ \midrule
R49	&	0,64935	\\ \midrule
R52	&	0,64935	\\ \midrule
R68	&	0,64935	\\ \midrule
R71	&	0,64935	\\ \midrule
R74	&	0,64935	\\ \midrule
R75	&	0,64935	\\ \midrule
R76	&	0,64935	\\ \midrule
R77	&	0,64935	\\ \midrule
R89	&	0,64935	\\ \midrule
R106&	0,64935	\\ \midrule
R107	&	0,64935	\\ \midrule
R108	&	0,64935	\\ \midrule
R116	&	0,64935	\\ \midrule
R117	&	0,64935	\\ \midrule
R133	&	0,64935	\\ \midrule
R67	&	0,58442	\\ \midrule
R69	&	0,58442	\\ \midrule
R70	&	0,58442	\\ \midrule
R72	&	0,58442	\\ \midrule
R73	&	0,58442	\\ \midrule
R109	&	0,58442	\\ \midrule
R110	&	0,58442	\\ \midrule
R130	&	0,58442	\\ \midrule
R131	&	0,58442	\\ \midrule
R31	&	0,51948	\\ \midrule
R51	&	0,51948	\\ \midrule
R54	&	0,51948	\\ \midrule
R57	&	0,51948	\\ \midrule
R59	&	0,51948	\\ \midrule
R64	&	0,51948	\\ \midrule
R66	&	0,51948	\\ \midrule
R78	&	0,51948	\\ \midrule
R79	&	0,51948	\\ \midrule
R80	&	0,51948	\\ \midrule
R81	&	0,51948	\\ \midrule
R82	&	0,51948	\\ \midrule
R83	&	0,51948	\\ \midrule
R84	&	0,51948	\\ \midrule
R85	&	0,51948	\\ \midrule
R111	&	0,51948	\\ \midrule
R132	&	0,51948	\\ \midrule
R141	&	0,51948	\\ \midrule
R142	&	0,51948	\\ \midrule
R30	&	0,45455	\\ \midrule
R33	&	0,45455	\\ \midrule
R35	&	0,45455	\\ \midrule
R36	&	0,45455	\\ \midrule
R53	&	0,45455	\\ \midrule
R55	&	0,45455	\\ \midrule
R58	&	0,45455	\\ \midrule
R60	&	0,45455	\\ \midrule
R61	&	0,45455	\\ \midrule
R62	&	0,45455	\\ \midrule
R65	&	0,45455	\\ \midrule
R86	&	0,45455	\\ \midrule
R87	&	0,45455	\\ \midrule
R88	&	0,45455	\\ \midrule
R90	&	0,45455	\\ \midrule
R124	&	0,45455	\\ \midrule
R125	&	0,45455	\\ \midrule
R126	&	0,45455	\\ \midrule
R127	&	0,45455	\\ \midrule
R128	&	0,45455	\\ \midrule
R129	&	0,45455	\\ \midrule
R134	&	0,45455	\\ \midrule
R34	&	0,38961	\\ \midrule
R37	&	0,38961	\\ \midrule
R38	&	0,38961	\\ \midrule
R112	&	0,38961	\\ \midrule
R118	&	0,38961	\\ \midrule
R140	&	0,38961	\\ \midrule
R32	&	0,32468	\\ \midrule
R39	&	0,32468	\\ \midrule
R40	&	0,32468	\\ \midrule
R41	&	0,32468	\\ \midrule
R42	&	0,32468	\\ \midrule
R43	&	0,32468	\\ \midrule
R50	&	0,32468	\\ \midrule
R105	&	0,32468	\\ \midrule
R120	&	0,32468	\\ \midrule
R121	&	0,32468	\\ \midrule
R122	&	0,32468	\\ \midrule
R135	&	0,32468	\\ \midrule
R146	&	0,32468	\\ \midrule
R48	&	0,25974	\\ \midrule
R114	&	0,25974	\\ \midrule
R119	&	0,25974	\\ \midrule
R143	&	0,25974	\\ \midrule
R144	&	0,25974	\\ \midrule
R145	&	0,25974	\\ \midrule
R44	&	0,19481	\\ \midrule
R45	&	0,19481	\\ \midrule
R46	&	0,19481	\\ \midrule
R47	&	0,19481	\\ \midrule
R104	&	0,19481	\\ \midrule
R113	&	0,19481	\\ \midrule
R115	&	0,19481	\\ \midrule
R123	&	0,19481	\\ \midrule
R91	&	0,12987	\\ \midrule
R92	&	0,12987	\\ \midrule
R93	&	0,12987	\\ \midrule
R94	&	0,12987	\\ \midrule
R95	&	0,12987	\\ \midrule
R96	&	0,12987	\\ \midrule
R97	&	0,12987	\\ \midrule
R98	&	0,12987	\\ \midrule
R99	&	0,12987	\\ \midrule
R136	&	0,12987	\\ \midrule
R137	&	0,12987	\\ \midrule
R138	&	0,12987	\\ \midrule
R139	&	0,12987	\\ \midrule
R56	&	0,064935	\\ \midrule
R63	&	0,064935	\\ \midrule
R100	&	0,064935	\\ \midrule
R101	&	0,064935	\\ \midrule
R102	&	0,064935	\\ \midrule
R103	&	0,064935	\\

\bottomrule    
\\
\caption{Reletionships ordered by Degree Prestige Centrality}
\label{tab:RelInDC}

\end{longtable}


\end{document}